# ODMRP with Quality of Service and local recovery with security Support

Farzane kabudvand
Computer Engineering Department zanjan
Azad University
Zanjan, Iran
E-mail: fakabudvand@yahoo.com

**Abstract**
In this paper we focus on one critical issue in mobile ad hoc networks that is multicast routing and propose a mesh based "on demand" multicast routing protocol for Ad-Hoc networks with QoS (quality of service) support.
Then a model was presented which is used for create a local recovering mechanism in order to joining the nodes to multi sectional groups at the minimized time and method for security in this protocol we present .

## 1.Introduction

*Multicasting is the transmission of packet to group of hosts identified by destination address.*

A multicast datagram is typically delivered to all members of its destination host group with the same reliability as regular unicast datagrams[4]. In the case of IP, for example, the datagram is not guaranteed to arrive intact at all members of the destination group or in the same order relative to other datagrams.

Multicasting is intended for group-oriented computing. There are more and more applications in which one-to-many dissemination is necessary. The multicast service is critical in applications characterized by the close collaboration of teams (e.g., rescue patrols, military battalions, scientists, etc.) with requirements for audio and video conferencing and sharing of text and images [3].

A MANET consist of a dynamic collection of nodes without the aid of the infrastructure of centralized administration . the network topology can change randomly and rapidly at predictable times.

The goal of MANETs is to extend mobility into the realm of autonomous, mobile, wireless domains, where a set of nodes form the network routing infrastructure in an ad hoc fashion. The majority of applications for the MANET technology are in areas where rapid deployment and dynamic reconfiguration are necessary and the wire-line network is not available [4]. These include military battlefields, emergency search and rescue sites, classrooms, and conventions where participants share information dynamically using their mobile devices.

QoS (Quality of Service) routing is another critical issue in MANETs. QoS defines nonfunctional characteristics of a system that affect the perceived quality of the result. In multimedia, this might include picture quality, image quality, delay, and speed of response. From a technological point of view, QoS characteristics may include timeliness (e.g., delay or response time), bandwidth (e.g., bandwidth required or available), and reliability (e.g., normal operation time between failures or down time from failure to restarting normal operation) [8].

In this paper, we propose a new technique for supporting QoS Routing for this protocol, and a technique then a model was presented which is used to create a local recovering mechanism in order to joining the nodes to multi-sectional groups at the minimized time, the fact that increases reliability of the network and prevents data wastage while distributing in the network.

## 2.Proposed protocol Mechanism
### A. Motivation

ODMRP[1] provides a high packet delivery ratio even at high mobility, but at the expense of heavy control overhead. It does not scale well as the number of senders and traffic load increases. Since every source periodically floods advertising RREQ[2] packets through the network, congestion is likely to occur when the number of sources is high . So control overhead is one of the main weaknesses of ODMRP, under the presence of multiple sources, CQMP solved the this problem, but both of these protocols have common weakness which is the lack of any admission control policy and resource reservation mechanism. Hence, to reduce the overhead generated by the control packets during the route discovery and apply admission control to network traffics, proposed protocol adopts two efficient optimization mechanisms. One is applied on nodes that cannot support QoS requirements, thus ignore the RREQ packet. The other is for every intermediate node and based on the comparison of available bandwidth of each node versus required bandwidth according to node position and neighboring node's role (sender, intermediate, receiver …). To address control packet problem, we use CQMP protocol's idea in RREQ packet consolidation, moreover we apply an admission control policy along with bandwidth reservation to our new protocol.

---
[1] On-demand Multicast Routing Protocol
[2] route request packet

### B. Neighborhood maintenance

Neighborhood information is important in proposed protocol.. To maintain the neighborhood information, each node is required to periodically disseminate a "Hello" packet to announce its existence and traffic information to its neighbor set. This packet contains the $B_{available}$ of the originator and is sent at a default rate of one packet per three seconds with time to live (TTL) set to 1. Every node in the network receives the Hello packet from its neighbors, maintains a neighbors list that contains all its neighbors with their corresponding traffic and co-neighbor number.

### C. Route discovery and resource reservation

Proposed protocol conforms to a pure on-demand routing protocol. It neither maintains any routing table, nor exchange routing information periodically. When a source node needs to establish a route to another node, with respect to a specific QoS requirement, it disseminates a RREQ that includes mainly, the requested bandwidth, delay and node's neighbor list. Hence, each intermediate node, upon receiving the RREQ performs the following tasks;

- Updates its neighbor's co-neighbor number;
- Determines whether it can consolidate into this RREQ packet information about other sources from which it is expecting to hear a RREQ. When a source receives a RREQ from another source, it processes the packet just as non-source intermediate node does, in addition checks its INT to determine if it would expire within a certain period of time, in other words the source checks if it is about to create and transmit its own RREQ between now and TIME-INTERVAL. If so, it adds one more row to the RREQ.
- Tries to respond to QoS requirements by applying a bandwidth decision in reserving the requested bandwidth $B$ which described in the follow, and before transmitting the packet appends its one-hop neighbor list along with their corresponding co-neighbor number to the packet.

As the RREQ may contain more than one Source Row, the processing node goes through each and every Source Row entry in the RREQ, and make admission decision for non-duplicated rows.

Admission decision is made at the processing node and it's neighbors listed in neighbor table as described in Section 3. If the request is accepted and there was enough bandwidth, the node will add a route entry in its routing table with status *explored*. The node will remain in explored status for a short period of $T_{explored}$. If no reply arrives at the explored node in time, the route entry will be discarded at the node and late coming reply packets will be ignored. Thus, we reduce the control overhead as well as exclude invalid information from the node's routing table.

Upon receiving each request packet, as the RREQ may contain more than one Source Row the receiver goes through each entry in the packet, builds and transmits a REPLY packet based upon matched entries along the reverse route. Available bandwidth of intermediate and neighboring nodes may have been changed due to the activities of other sessions. Therefore, similar to the admission control in RREQs, upon receiving a RREP, nodes double check the available bandwidth to prevent possible changes during the route discovery process. If the packet is accepted, the node will update the route status to *registered*. After registration, the nodes are ready to accept the real data packets of the flow. The node will only stay in registered status for a short period of $T_{registered}$. If no data packet arrives at the registered node in time, it means that the route was not chosen by the source. Then the route entry will be deleted at the node.

When any node receives a REPLY packet, it checks if the next node Id in any of the entries in the REPLY matches its own. If so, it realizes that it is on the way to a source, It checks its own available bandwidth and compares it with required bandwidth of this flow, then checks its one-hop neighbor's available bandwidth which recorded in the neighbor table. If there was enough bandwidth it sets a flag indicating that it is part of the FORWARDING GROUP for that multicast group, and then builds and broadcasts its own REPLY packet.

When a REPLY reaches a source, a route is established from the source to the receiver. The source can now transmit data packets towards the receiver. A Forwarding Group node will forward any data packets received from a member for that group.

### D. Data Forwarding

After constructing the routes, the source can send packets to multicast group via selected routes and forwarding nodes. Upon receiving a data packet forwards it, only when;

It is not a duplicate packet, Forwarding flag for this session has not expired,There was an entry with registered or reserved status corresponds to this session.

It then changes its *'registered'* status to *'reserved'*. The node will only stay in *reserved* status for a short period of T_reserved. This procedure minimizes the traffic overhead and prevents sending packets through the stale routes.

## 3. Local Recovery Mechanism based on proposed protocol with reliability

In this section the mechanism of local recovery will be discussed on the basis of a suggested protocol. The suggested method leads to fast improvement of the network and therefore the destination can be connected to the source through a new route. Discovered routes between destination and source may be corrupted for many reasons most of which could be occurred because of removing in nodes.

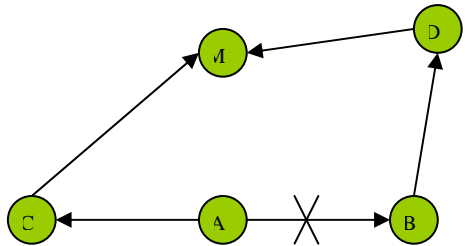

**Fig3. local recovery with proposed method**

By considering figure .3, if direct link A-B corrupts an indirect route of A to B will be formed by C which stands next door to them. In this condition if some package with many steps is sent to find next node regenerating of the present route will be possible and there is no need to regenerate the end to end by three times. Algorithm follows as that when a middle node FG recognizes route corruption between itself and the next step it places data on its buffer and starts to set a timer. Then it sends the package with more steps (i.e. two steps) and puts on it a set of nodes which are placed at a farther space between source and destination. Receiving this package, every node begins to consider whether its name to be there or not. If the address of the node corresponds to one of the current addresses, the answer package may be sent and as a result of that it can be sent through a new route. But if the answer isn't received by the end of given time determined on the timer, the package is thrown away and another route may be discovered again. Every node which receives a local regained package and its answer will function as a FG for that destination. Thus every node should be aware of its FG between itself and destination. In this way we can recognize some alteration in the structure of the protocol, that is, every FG add only the name of the node to the received answer package by sending it up to a higher node. In other words the existing addresses in the answer package are not to be omitted rather some desired address of FG node is added to the answer package. In this way every FG can be aware of other FG between itself and destination, and starts to use them. Here the number of the steps is considered 2.

As it can be seen in figure 4. while sending the answer package of membership destination in this method puts the address of the proceeding group in the package and sends it. Now FGs also do the same. Therefore every node can recognize the member of the proceeding group of all proceeding nodes between itself and destination and begin to send local recovering package in case of route corruption.

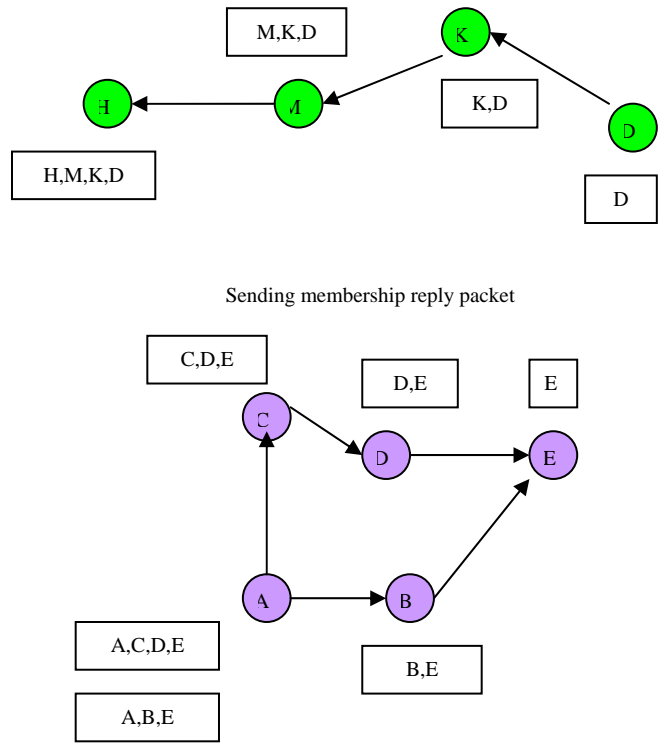

Sending membership reply packet

Fig4.sending local recovery packet and update address

Given timer is taken 1 second, namely if FG which sends data fails to receive the same data after utmost one second from following FG, then it discovers a route corruption and modulates another timer amount to 0/1 sec. in order to receive the answer package and therefore a new route can be resulted. During this the package is put in a temporary buffer. If new routes cannot be found the package would be thrown away.

## 4. Security in mobile Ad hoc networks

In mobile Ad hoc networks , due to unreliable data and lack of infrastructure , providing secure communications is a big challenge . in wired and wireless networks cryptographic techniques are used for secure communications.

The key is a piece of input information for cryptography. If the key is discovered ,the encrypted information can be revealed.

There are some domaining trust model because the authentication of key ownership is important .

One of important models is centralized .In this model we can use a hierarchical trust structure .

It is necessary for security we distribute the control trust to multiple entities that is the system public key is distributed to whole network . because a single certification node could be a security bottleneck and multiple replicas of certification node are fault tolerant.

In proposed technique for security we consider number of nodes that they hold a system private key share and are able of producing certificates. These nodes are named s-node . s-nodes and forwarding nodes( a sunset of non s-node) generate a group .

When a s-node enters the network it broadcasts a request packet . this packet has extra attributes this packet consist of TTL field , this field decrease by 1 as the packet leaves the node .

When a node receives the request packet it first checks the validity of packet before taking any further actions. Then discards non-authenticated packets. Neighbor nodes to s-nodes receive the request and rebroadcast it .This process continues at other nodes .

When another s-node receives the packet from neighbor ( example node B ) it could send back a server reply message to neighbor ( example node B ).

When B receives the join reply packet , it learns that it's neighbor is a s-node and it is on the selected path between two server and set the forwarding attribute to 1 .

After all s-nodes finish the join procedure the group mesh structure is formed .

This procedure can create security in whole network .

## 5 .Performance Evaluation

We implement the proposed protocol in GloMoSim. The performance of the proposed scheme is evaluated in terms of average Number of RREQ sent by every node, end-to-end delay, and packet delivery ratio. In the simulation, we modeled a network of 50 mobile hosts placed randomly within a $1000*1000\, m^2$ area. Radio propagation range for each node was 250 meters and channel capacity was 2Mbit/sec. Each simulation runs for 300 seconds of simulation time. The MAC protocol used in our simulations is IEEE 802.11 DCF [22]. We used Constant Bit Rate as our traffic. The size of data payload was 512 bytes. The nodes are placed randomly within this region. The multicast sources are selected from all 50 nodes randomly and most of them act as receivers at the same time. The mobility model used is random waypoint, in which each node independently picks a random destination and speed from an interval (*min*, *max*) and moves toward the chosen destination at this speed. Once it reaches the destination, it pauses for *pause* number of seconds and repeats the process. Our *min* speed is 1 m/s, *max* speed is 20 m/s and *pause* interval is 0 seconds. The RREQ interval is set at 3 second. The HELLO refresh interval is the same as the RREQ interval. We've varied the following items: mobility speed, number of multicast senders and network traffic load.

*Performance Metrics used:*
- **RREQ Control Packet Load**: The average number of RREQ packet transmissions by a node in the network.
- **Packet delivery Ratio***:* The ratio of data packets sent by all the sources that is received by a receiver.
- **End to end delay**: refers to the time taken for a packet to be transmitted across a network from source to destination.

## 6.Results

In Fig. 5, we calculated the delivery ratio of data packets received by destination nodes over data packets sent by source nodes. Without admission control, more packets are injected into the network despite they cannot reach destinations. These packets waste a lot of channel bandwidth. On the other hand, if the admission control scheme is enabled, the inefficiency usage of channel resource can be limited and the saturation condition can be alleviated. Since proposed protocol has less RREQ packet transmissions than ODMRP and CQMP, there is less chance of data packet loss by collision or congestion. Owning to additional Hello overhead, proposed protocol performs a litter worse when there are few sources. The data delivery ratio of evaluated protocols decreases as the number of sources increases under high mobility conditions, but proposed protocol constantly maintains about 4 to 5 percent higher packet delivery ratio than others because of reduction of join query overhead.

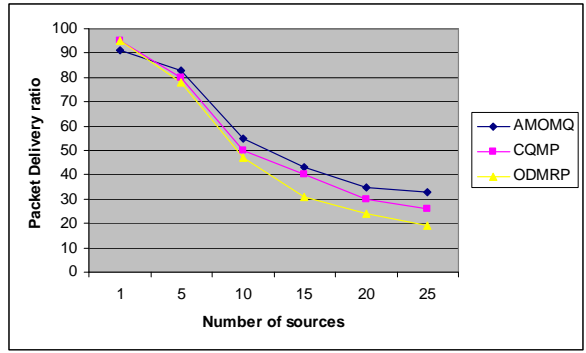

Fig5 Packet Delivery Ratio as a function of Number of Sources

## 7. Conclusion

In this paper, we have proposed a mesh-based, on-demand multicast routing protocol with admission control decision, proposed protocol, which similar to CQMP uses consolidation of multicast group membership advertising packets plus admission control policy.

then model was presented which is used to create a local recovering mechanism in order to joining the nodes to multi-sectional groups at the minimized time, the fact that increases reliability of the network and prevents data wastage while distributing in the network. In this mechanism a new package known as local recovering package was created by using of a membership suit package and placing the address of the nodes between a proceeding group and destination. Here we considered the number of steps restricted but it can be changed. We implemented proposed protocol using GlomoSim and show by simulations that proposed protocol shows up to 30 percent reduction in control packet load. In addition, our results show that as the number of mobile sources increased and under large traffic load, proposed protocol performs better than ODMRP and CQMP in terms of data packet delivery ratio, end-to-end delay and number of RREQ packets. By proposed scheme, network saturation under overloaded traffic can be alleviated, and thereby, the quality of service can be improved.